\documentclass[11pt,a4paper]{article}

\usepackage[T1]{fontenc}
\usepackage[utf8]{inputenc}
\usepackage{graphicx}      
\usepackage{amsmath,amssymb}
\usepackage{geometry}
\usepackage{caption}
\usepackage{setspace}
\usepackage{tikz}
\usepackage{subcaption}
\usetikzlibrary{positioning}
\usepackage{booktabs}
\usepackage{siunitx}   
\usepackage[T1]{fontenc}
\usepackage[scaled=0.92]{helvet}

\usepackage{tikz}
\usepackage{url} 
\usetikzlibrary{positioning,fit,backgrounds,arrows.meta,calc}
\usetikzlibrary{arrows.meta,shapes.geometric}
 
\definecolor{cRoot}{HTML}{4D4D4D}
\definecolor{cMeta}{HTML}{0072B2}
\definecolor{cHead}{HTML}{009E73}
\definecolor{cArc}{HTML}{D55E00}
\definecolor{cBody}{HTML}{CC79A7}

\definecolor{oiBlue}{HTML}{0072B2}
\definecolor{oiVermillion}{HTML}{D55E00}
\definecolor{oiGreen}{HTML}{009E73}
\definecolor{oiOrange}{HTML}{E69F00}
\definecolor{oiPurple}{HTML}{CC79A7}
\definecolor{oiSky}{HTML}{56B4E9}
\definecolor{ink}{HTML}{333333}   

\newcommand{\id}[1]{{\ttfamily\fontsize{6}{7}\selectfont #1}}
\newcommand{\un}[1]{{\color{black!60}[#1]}}
 
\tikzset{
  grp/.style={draw=#1, fill=#1!10, rounded corners=2pt, thick,
              font=\bfseries\fontsize{7}{8}\selectfont, inner sep=3pt,
              align=center, text=black},
  bx/.style={draw=#1, fill=white, rounded corners=1.5pt, semithick,
              font=\fontsize{6}{7.4}\selectfont, inner sep=3pt,
              align=left, anchor=north west},
  lnk/.style={-{Stealth[length=4pt]}, thick, draw=black!60},
  ttl/.style={font=\bfseries\fontsize{6.5}{7.5}\selectfont, text=#1},
}

\tikzset{
    mainbox/.style={
        rectangle,
        rounded corners=4pt,
        draw=LuGREborder,
        line width=0.8pt,
        align=center,
        font=\sffamily,
        inner sep=0pt
    },
    whitebox/.style={
        mainbox,
        fill=white
    },
    lightbox/.style={
        mainbox,
        fill=LuGRElight,
        draw=none
    },
    darkbox/.style={
        rectangle,
        rounded corners=4pt,
        fill=LuGREblue,
        draw=none,
        text=white,
        align=left,
        font=\sffamily,
        inner sep=7pt
    },
    groupbox/.style={
        rectangle,
        rounded corners=13pt,
        draw=LuGREborder,
        line width=0.8pt,
        fill=white,
        inner sep=0pt
    },
    flowarrow/.style={
        -{Latex[length=3.8mm,width=2.1mm]},
        line width=0.85pt,
        draw=LuGREblue
    },
    dashedflow/.style={
        -{Latex[length=3.8mm,width=2.1mm]},
        line width=0.85pt,
        draw=LuGREblue,
        dashed
    },
    sectionlabel/.style={
        font=\sffamily\bfseries\scriptsize,
        text=LuGREblue
    }
}

\definecolor{teal_fa}{HTML}{DC143C}
\definecolor{blue}{HTML}{0000FF}
\definecolor{boxgrey}{HTML}{404040}
\definecolor{LuGREblue}{RGB}{18,48,78}
\definecolor{LuGRElight}{RGB}{228,232,238}
\definecolor{LuGREborder}{RGB}{202,205,209}
\definecolor{LuGREtext}{RGB}{65,75,85}

\title{\vspace{-1cm}A catalog of dispersive ionospheric and plasmaspheric arcs from cislunar and lunar-surface GNSS radio occultations}

\author{
\footnote{F. Fiorina is the corresponding author (francesco.fiorina@polito.it).}\underline{Francesco Fiorina}$^{1,*}$, Alex Minetto$^{1,*}$,Rebecca Ghidoni $^{2,\dag}$, Marco Guerra $^{2,\dag}$, \\ Claudio Cesaroni $^{2,\dag}$, Luca Spogli $^{2,\dag}$, Andrea Nardin $^{1,*}$, Fabio Dovis$^{1,*}$
}

\date{}

\begin{document}
\maketitle
\thispagestyle{empty}

\noindent
$^{1}$Department of Electronics and Telecommunications (DET)/Politecnico di Torino, Turin, Italy\\
$^{2}$Istituto Nazionale di Geofisica e Vulcanologia (INGV), Rome, Italy\\
$^{*}$name.surname@polito.it
$^{\dag}$name.surname@ingv.it

\vspace{0.5cm}

\section*{Abstract}
This dataset is a catalog of 263 dispersive arcs induced by ionospheric and plasmaspheric occultation observed in GNSS signals recorded by the Lunar GNSS Receiver Experiment (LuGRE) payload during Blue Ghost Mission 1 (BGM1), in cislunar space and on the lunar surface. Occultation arcs are identified and classified as ingress, egress or tangent by combining independently GPS and Galileo dual-frequency observables and computing the rate of the total electron content (ROT) and ROT index (ROTI), and are validated through a geometric criterion based on the grazing altitude of the tangent point. No distinction is made between the ionospheric and plasmaspheric contributions to each arc, which is treated as a single dispersive signature. For each arc, the catalog provides timing, grazing altitude, summary ROT/ROTI/$\Delta$STEC attributes, the geomagnetic activity level, and time series of leveled slant total electron content (STEC), not corrected for instrumental biases, ROT, ROTI, tangent-point position and local time. The HDF5 (Hierarchical Data Format version 5) file contains a header with dataset-wide metadata and per-class statistics, and one entry per arc. It enables rapid selection of occultation arcs for statistical studies and targeted analyses, and complements ground-based and low Earth orbit radio-occultation data during the declining phase of solar cycle 25.

\vspace{0.5cm}

\section*{Background \& Summary}
Dispersive occultation arcs --- transient, frequency-dependent perturbations imprinted on Global Navigation Satellite Systems (GNSS) signal observables during radio occultation --- are a signature of the near-Earth's plasma environment, mostly composed by ionospheric and plasmaspheric plasma.
The ionosphere (about 80--1500~km in altitude$^{1}$
) and the plasmasphere, which extends it along closed geomagnetic field lines, are coupled ionized regions connected through the so-called transition region$^{2}$.
Both regions are dispersive media for GNSS signals$^{1,3}$. 
Their plasma frequencies lie far below the L-band carriers, so GNSS signals cross them without reflection, but accumulate a group delay and a phase advance that depend on the electron density along the ray path, referred to as total electron content (TEC)$^{3,4}$. 
The plasmaspheric contribution to the overall TEC is not negligible: at mid-latitudes during solar minimum, the plasmaspheric electron content has been estimated at about 25--45\% of the GPS TEC during daytime and above 50--60\% at night$^{5}$. 
However, dual-frequency observations are sensitive only to the electron content integrated along the whole ray path, and therefore cannot separate the two contributions$^{1}$.
\\\\
A systematic characterisation of the ionosphere and plasmasphere requires long-duration observations spanning a wide range of grazing altitudes and geometries. Despite their continuous operations, radio occultation missions in low Earth orbit (LEO), geostationary orbit (GEO) and ground sounding experiments sample only a limited portion of this range$^{1,6}$.
Such observations are relevant to space weather and ionospheric monitoring and modelling, and to the correction models used to compensate for excess delay affecting GNSS observables$^{7}$.\\\\
In these contexts, the Lunar GNSS Receiver Experiment (LuGRE), flown on Blue Ghost Mission 1 (BGM1), offered a complementary vantage point: across the commissioning, transit, low lunar orbit, and lunar-surface phases of the mission, its payload recorded GNSS signals with tangent-point altitudes spanning 0--18\,000~km. Therefore, LuGRE enabled a single instrument to observe dispersive effects on signals propagating through both the ionosphere and the plasmasphere, from a unique and previously unavailable point of view.\\\\ LuGRE operated between 15 January and 16 March 2025, at the beginning of the declining phase of Solar Cycle~25, under quiet solar and geomagnetic activity, as indicated by the F10.7 solar flux$^{8,9}$ and the planetary Kp index$^{10,11}$, respectively. 
The peculiar observation geometry of LuGRE is depicted in Figure~1
, which shows the lines of sight connecting the transmitting antenna pattern main and sidelobes to the receiver antenna onboard BGM1. Different mission phases and observations were hence characterized by different dispersive effects, depending on the path length through the ionosphere and plasmasphere, as well as by diverse sun illumination conditions of these layers.\\\\
The majority of the acquired and tracked GNSS signals were received through the main lobes of the transmit antenna patterns, since the signal paths from GNSS satellites located behind the Earth, as seen from the receiver, grazed the Earth limb$^{12}$. Therefore, for a considerable amount of time those signals produced altered GNSS observables, carrying information about the dispersive media. 
\begin{figure*}[t]
    \centering
        \includegraphics[trim={0.1cm 2cm 0.1cm 4cm},clip,width=1\linewidth]{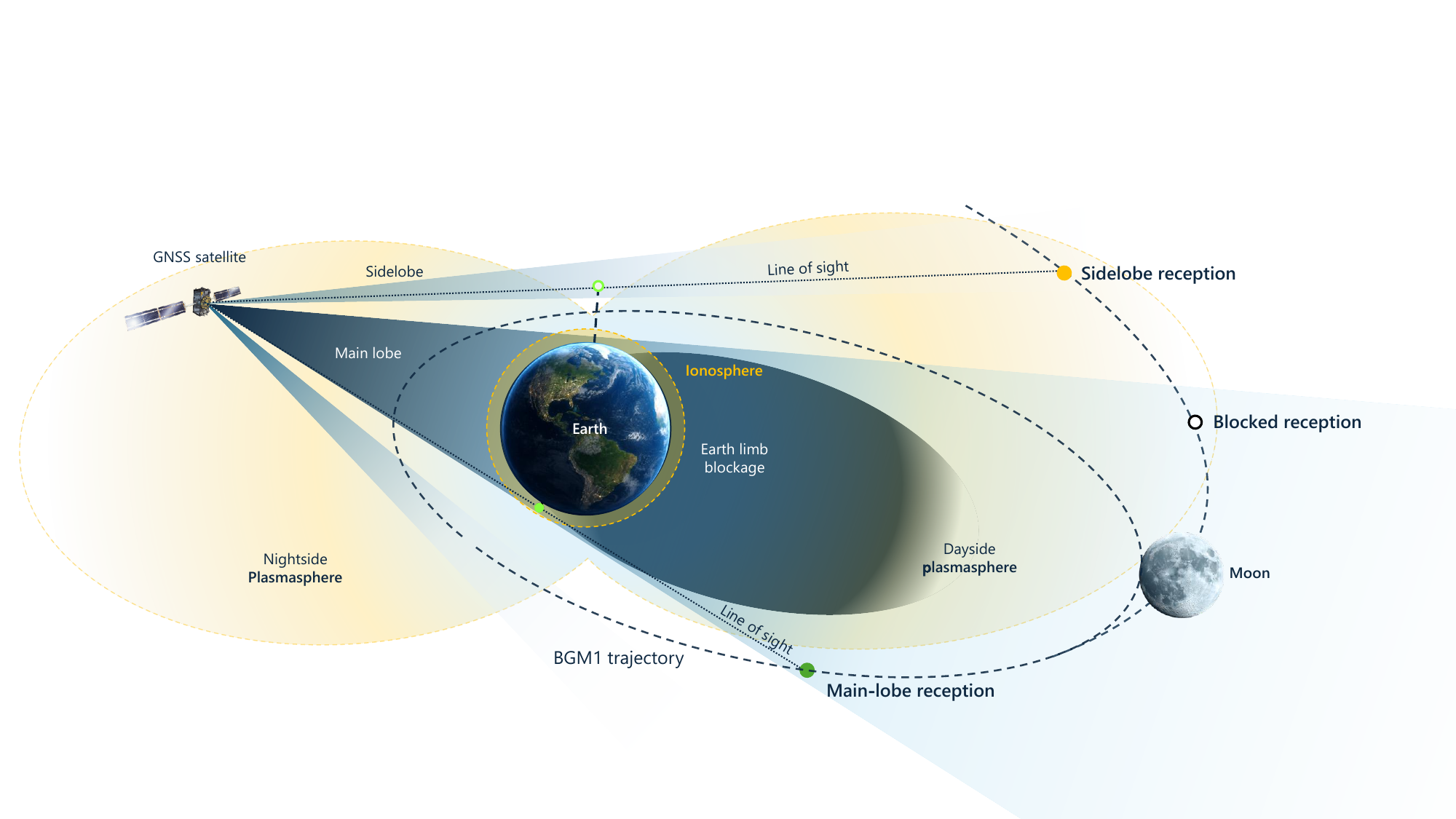}
        \caption{\textbf{Observation geometry of GNSS signals received by LuGRE along the BGM1 trajectory}. Schematic (not to scale) of the Lunar GNSS Receiver Experiment (LuGRE) on the Blue Ghost Mission 1 (BGM1) trajectory (dashed line) and of the Global Navigation Satellite System (GNSS) signals it receives. Signals are transmitted through the main lobe (dark blue) or sidelobes (light blue) of the satellite antenna and can be received along the trajectory (green dot: main-lobe reception; yellow dot: sidelobe reception) unless the line of sight is obstructed by the Earth (Earth limb blockage; open circle: blocked reception). Small green dots on the lines of sight mark the tangent points, i.e. the points of closest approach to the Earth. Lines of sight whose tangent point lies within the ionosphere (yellow ring) or the nightside/dayside plasmasphere (yellow shaded regions) accumulate dispersive delay and are the subject of this dataset.}
    \label{fig:LuGRE_POV}
\end{figure*}
Although LuGRE data can support several investigations, the vantage point offered by BGM1 makes them particularly valuable for space-climatology studies. However, the public LuGRE dataset$^{13}$ was not designed to isolate the observations relevant to this purpose. \\\\
To address this need, we developed an algorithm that detects, characterizes and classifies the occultation arcs in the LuGRE GNSS observables across all four mission phases. Since dual-frequency observations cannot separate the ionospheric and plasmaspheric contributions, each arc is treated as the signature of the whole ionosphere--plasmasphere system along the signal path. Here, “occultation” is used in the radio-occultation sense: an arc is a period in which the tangent point of the signal path sweeps through the ionospheric and plasmaspheric layers, probing them along the ray path. Ingress and egress arcs end or begin with Earth blockage of the signal, whereas tangent arcs graze the dispersive layers without Earth blockage.
Here we describe the resulting catalog, which comprises 263 arcs derived from an extended version of a preliminary algorithm limited to main-lobe arcs elaboration$^{14}$. 
A single HDF5 file provides, for each arc, the onset and end times, the Kp index and G-scale labels according to the NOAA scale of geomagnetic storms$^{10}$, 
the grazing altitudes of the tangent point and TEC-derived indices, i.e., the Rate Of TEC (ROT) and ROT Index (ROTI), together with the corresponding pre-processed time series. This arc-level representation removes the need to repeat the low-level processing of the original mission data: users can select arcs of interest from their summary metadata, for example arcs with possible TEC gradient irregularity signatures or arcs whose tangent point grazing altitudes exceed a given threshold, and retrieve only the associated time series. The catalog is therefore intended to support both statistical studies of ionospheric and plasmaspheric effects observed from cislunar space and targeted analyses of individual occultation arcs, including the propagation impairments affecting GNSS signals received at lunar distances.


\section*{Methods}
\label{sec:Methods}
The presented dataset results from the post-processing of the \texttt{Lunar GNSS Receiver Experiment (LuGRE) Mission Data} public dataset$^{13}$. 
This catalog contains a characterization of each arc observed by the LuGRE payload. Each arc is classified by inspecting specific TEC-derived parameters, namely the ROT and ROTI. Differently from a pure geometrical approach, relying on TEC-derived parameters gives the possibility of identifying the exact timespan in which the dispersive effects are impactful. To this end, the GNSS raw observables are processed through a dedicated pipeline summarized in Figure~2. 

The chain consists of the parsing of the mission data, a pre-processing stage, the computation of the TEC-derived indices, and the identification and classification of the arcs; each step is detailed below.
\begin{figure*}[t]
    \centering
        \includegraphics[trim={2cm 4cm 2cm 6cm},clip,width=1\linewidth]{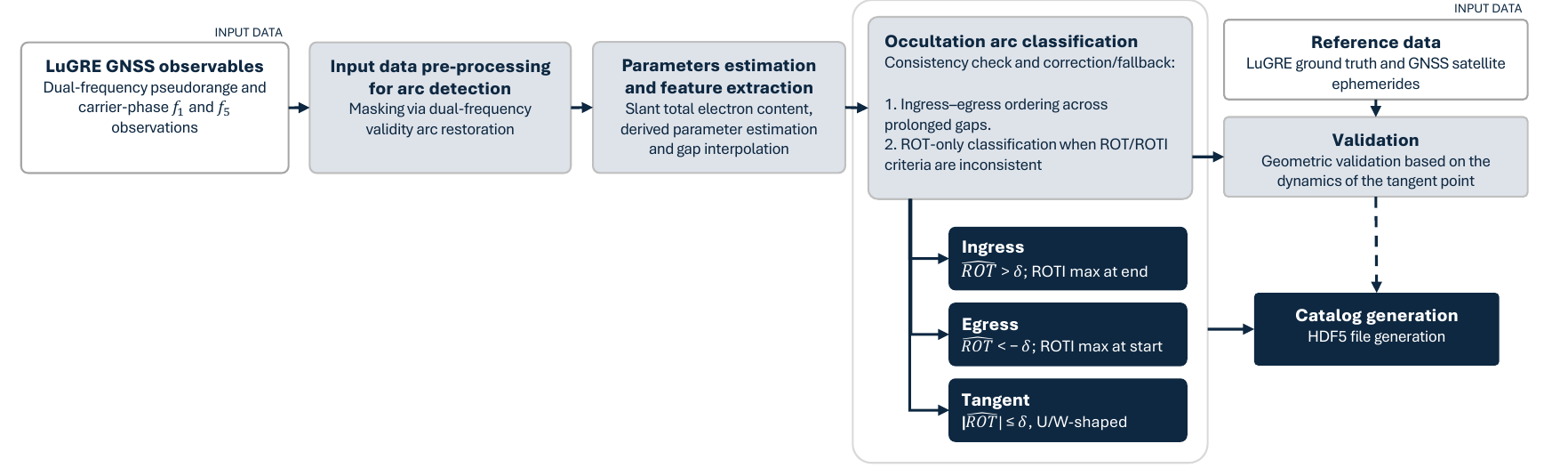}
        \caption{\textbf{Workflow for the generation and validation of the catalog of dispersive occultation arcs.} Processing chain from the dual-frequency Global Navigation Satellite System (GNSS) observables recorded by the Lunar GNSS Receiver Experiment (LuGRE) to the catalog released as HDF5 files. Outlined boxes labelled “Input data”: LuGRE observables and reference data (LuGRE ground truth and GNSS satellite ephemerides). Grey boxes: processing steps, from arc detection to slant total electron content (STEC) estimation, classification and geometric validation. Dark blue boxes: output arc classes, assigned from the sign and magnitude of the rate of change of TEC (ROT) relative to the threshold $\delta$, and from the position of the maximum of the ROT index (ROTI) within the arc. Ingress arcs satisfy $ROT \ge \delta$  with the ROTI maximum at the end of the arc, egress arcs $ROT \le -\delta$ with the ROTI maximum at its start, and tangent arcs $|ROT| \leq \delta$ with a U/W-shaped time series. Consistency checks and fallback rules resolve cases where the two criteria disagree. Solid arrows: data flow. Dashed arrow: geometric validation, based on the dynamics of the tangent point, identifies which of the classified arcs are also geometrically consistent.}
    \label{fig:workflow}
\end{figure*}
\subsection*{Input data}
The catalog is derived from three publicly available sources. The primary input is the \texttt{Lunar GNSS Receiver Experiment (LuGRE) Mission Data} dataset$^{13}$, 
which contains the GPS and Galileo observables recorded on the L1/E1 and L5/E5a bands at 1~Hz across all BGM1 mission phases, together with the table of the executed scientific operations pursued by the payload. LuGRE data were, in fact, collected during a series of discrete operations, i.e., observation windows in which the payload was powered on and pointed towards the Earth; in each operation, the receiver could run in real-time processing (RTP) mode, producing GNSS observables, and/or record raw in-phase and quadrature (I/Q) signal snapshots$^{12,13}$. 
Since the classification relies on dual-frequency observables, only operations that include RTP observables are considered. Of the 21 such operations, OP3\_0 was discarded because a single satellite signal was visible and tracked for less than 100~s, so that 20 operations, each processed independently, contribute to the catalog.\\\\
Two ancillary data sources are additionally used for the validation. The position of the BGM1 lander, and hence of the LuGRE receiver, is obtained at each observation epoch from the SPICE kernels of the NASA Commercial Lunar Payload Services (CLPS) program archived by the Navigation and Ancillary Information Facility (NAIF)$^{15}$ 
(task order TO19D, meta-kernel \texttt{clps\_to19d\_v001.tm}), expressed in the ITRF93 Earth-fixed frame. GNSS satellite orbits are computed from the daily multi-GNSS broadcast navigation files (RINEX~3) of the International GNSS Service (IGS)$^{3}$ 
, retrieved from the NASA Crustal Dynamics Data Information System (CDDIS) archive$^{16}$ 
for the start day of each operation. \\\\
The geomagnetic activity associated with each arc is derived from the definitive planetary Kp index distributed by GFZ Potsdam$^{17,11}$. 

\subsection*{LuGRE Data parsing}
The telemetry files of each RTP operation are decoded with the LuGRE Navigation Analysis and Reporting Tool (LuNART)$^{18}$, 
available at https://github.com/NavsasPolito/LuNART, which extracts the time series of $C/N_0$, pseudorange and carrier-phase observables for each tracked signal on a common 1~Hz epoch grid into a per-operation MATLAB structure used by the subsequent processing steps.

\subsection*{Input data pre-processing}
Before the TEC-derived indices are computed, the GNSS measurements undergo a pre-processing stage consisting of three steps: cycle-slip detection, signal-quality masking, and the construction of an arc-validity mask.

First, cycle slips are detected on the code-minus-carrier (CMC) time series of each signal through Grubbs outlier test$^{19}$. 
The epochs at which a cycle slip is detected are flagged to be properly treated in the subsequent processing steps, namely to re-initialize the leveling filter and to discard corrupted ROT estimates, as described in the next subsection.

Second, a masking procedure is applied to the $C/N_0$ estimates: all pseudorange and carrier-phase measurements whose $C/N_0$ falls below 20~dB-Hz are excluded, to prevent observations collected under degraded tracking conditions from introducing noise into the derived quantities.

Third, since tracking losses may occur at any point during a satellite pass, an arc-validity mask is built to identify the epochs suitable for arc detection and classification. The mask is built in two steps. The first step flags a satellite as visible at a given epoch if its signal is received on at least one of the two frequencies, $f_1$ or $f_5$: the satellite is regarded as "untracked" only when both are simultaneously absent, so that gaps occurring on a single frequency do not interrupt the visibility arc. The second step retains as valid only the epochs lying between the first and the last epoch at which both frequencies are simultaneously available, since the classification requires dual-frequency observables.
Within the validity timespan, gaps on a single frequency do not interrupt the arc, but they prevent the computation of the dual-frequency TEC-derived indices defined in the next subsection at the affected epochs. The resulting gaps in the parameter time series would be interpreted by the classification algorithm as arc boundaries, splitting a single occultation arc into several disjoint ones. To preserve arc continuity and integrity, the parameter series are interpolated across these gaps. Since interpolating at an early stage could introduce artefacts that would propagate through the processing chain, the interpolation is applied only at its end, to the ROTI series, as described in the next subsection.

\subsection*{Total electron content and derived parameters}
To obtain the parameters on which the identification and classification algorithm relies on, is at first necessary to compute the Slant Total Electron Content (STEC), i.e., the integral of the electron density along the ray path from the satellite to the receiver$^{3}$, 
commonly expressed in TEC units (1~TECU $= 10^{16}$~electrons/m$^2$). Signals transmitted at different carrier frequencies experience different group delays and phase advances; therefore, by exploiting the dispersive nature of the ionosphere and plasmasphere, STEC can be estimated$^{3,20}$. 
Dual-frequency pseudorange ($\rho$) and carrier-phase ($\varphi$, in meters) measurements collected from the same GNSS satellite can be combined to obtain the code-derived STEC
\begin{equation}
    \mathrm{STEC}_{\rho}=\frac{f_{1}^{2} f_{5}^{2}}{40.3\left(f_{1}^{2}-f_{5}^{2}\right)}\left(\rho_5 - \rho_1\right),
    \label{eq:STECpseudorange}
\end{equation}
and the phase-derived STEC
\begin{equation}
    \mathrm{STEC}_{\varphi}=\frac{f_{1}^{2} f_{5}^{2}}{40.3\left(f_{1}^{2}-f_{5}^{2}\right)}\left(\varphi_1 - \varphi_5\right),
    \label{eq:STECphase}
\end{equation}
where subscript $1$ refers to the L1/E1 band (central frequency 1575.42 MHz) and subscript $5$ to the L5/E5a (central frequency 1176.45 MHz) band. Eqs. (1)–(2) yield STEC in electrons/m$^2$; values are converted to TECU by multiplying by $10^{-16}$.
$\mathrm{STEC}_{\rho}$ provides an absolute but noisy estimate, whereas $\mathrm{STEC}_{\varphi}$ is considerably more precise but biased by the unresolved carrier-phase ambiguity.

Combining the absolute accuracy of the code-derived estimate with the precision of the phase-derived one, a recursive carrier-phase levelling algorithm$^{21}$ 
is adopted. The method is based on the code-minus-phase differencing of STEC estimates
\begin{equation}
    \Delta(t) = \mathrm{STEC}_{\rho}(t) - \mathrm{STEC}_{\varphi}(t).
\end{equation}
The offset between the two STEC estimates is tracked as a running mean:
\begin{equation}
    \Theta(t) = \frac{(K-1)\cdot\Theta(t-1) + \Delta(t)}{K},
\end{equation}
where $K$ is the count of valid epochs since the last filter reset. 
The leveled STEC is then obtained as
\begin{equation}
    \mathrm{STEC}_{\mathrm{leveled}}(t) = \mathrm{STEC}_{\varphi}(t) + \Theta(t).
\end{equation}
This filter needs to be reinitialized under four specific conditions: at the beginning of each dataset, at the first valid epoch, whenever a cycle-slip is flagged by the upstream receiver processing, or whenever the gap between consecutive valid samples exceeds a configurable threshold (e.g., 10~s).

The primary parameters used to classify the type of occultation arc, and more generally used to characterize the dispersive media such as the ionosphere and plasmasphere, are ROT and ROTI$^{22}$. 
In this methodology, the ROT, i.e., the first order time-derivative of STEC, is computed from the phase-derived STEC of Eq.~2
, which is less noisy than the code-derived and leveled estimates. 
To retrieve the ROT estimate, a three-point central finite-difference operator with second-order accuracy is computed on individual dual-frequency observables segments with no gaps exceeding one epoch:
\begin{equation}
    \mathrm{ROT}(t_k) = \frac{\mathrm{STEC}_{\varphi}(t_{k+1}) - \mathrm{STEC}_{\varphi}(t_{k-1})}{t_{k+1} - t_{k-1}}.
\end{equation}
expressed in TECU/s. At the segment endpoints, first-order forward and backward differences are adopted to minimize sample loss. The differentiation removes both the carrier-phase ambiguity and the instrumental biases, which can be assumed to be constant over consecutive epochs. Cycle slips, however, produce spurious spikes in the ROT that propagate to the ROTI$^{23}$; 
therefore, all ROT estimates within one epoch of a flagged cycle slip are discarded.

To describe the ROT variability, the ROTI can be computed, which corresponds to its standard deviation over a sliding window: 
\begin{equation}
    \mathrm{ROTI}_k = \sqrt{
        \frac{1}{N}
        \sum_{n=k-N/2}^{k+N/2-1}
        \left(\mathrm{ROT}_n - \overline{\mathrm{ROT}}\right)^2
    },
\end{equation}
where $\overline{\mathrm{ROT}}$ is the mean ROT over a window of $N$ samples.
The window length is adapted to the observable sampling rate: a 1-minute window is
typically adopted at 1~Hz sampling, or equivalently a 5-minute window at
30~s sampling$^{23,24}$. 

In this work, a 30-s window was selected to provide a sufficiently large number of ROT samples for a stable estimate of ROTI while limiting the spatial extent of the ionospheric and plasmaspheric region contributing to the estimate. ROTI is expressed in TECU/s (multiply by 60 for TECU/min, the unit commonly used in the literature). Note that it is computed here over a 30\,s window, shorter than the 1–5 min windows typically adopted when dealing with geodetic receivers with a sampling rate of $(1 s)^{-1} - (30 s)^{-1}$, so direct comparison with literature values should account for this difference.
In general, plasma irregularities and small-scale ionospheric perturbation can be diagnosed by inspecting the ROTI$^{3}$. 

Finally, the gaps in the ROTI series falling within the arc-validity mask are filled by Akima interpolation method$^{25}$ 
, yielding a restored ROTI series that is used only to preserve arc continuity during classification. The interpolated values are not used for any quantitative analysis and are not included in the catalog, which only provides the original ROTI series.

\subsection*{Arcs identification and classification}
Once the ROT and ROTI series have been computed, as described in the previous subsections, the dispersive arcs are identified and classified. The identification consists of three steps, i.e., segmentation, feature extraction, and grouping.

First, the restored ROTI series of each satellite is segmented into contiguous intervals, each representing a candidate arc. The restored series is used only to define the segment boundaries, whereas all features are computed on the original, non-interpolated ROT and ROTI series.

Second, the following features are extracted from each segment:
\begin{itemize}
    \item the minimum and maximum ROTI values and their temporal location within the segment, which indicate the epochs of significant ROT variation and are used to define the start and end of the arc;
    \item the segment duration, $L$;
    \item the time gap, $T_g$, to the subsequent segment of the same satellite. \end{itemize}

Third, consecutive segments are grouped to account for the case in which the ROTI series is characterized by several adjacent segments separated by short dropouts, caused either by tracking losses or by samples discarded by the ROTI windowing. Two consecutive segments are merged into a single candidate arc if $T_g < 300$~s; otherwise, they are treated as distinct arcs.

Finally, after grouping, the features are then recomputed on each candidate arc, together with the median ROT i.e., $\widehat{\mathrm{ROT}}$ obtained after removing outliers with Tukey's fences$^{26}$ 
. The median is preferred to the mean for its robustness to outliers, and provides the primary discriminant between arc types. To reduce the risk of misclassification, only candidate arcs with $L>100$\,s are classified. For instance, no arc could be classified for operation OP3\_0, since its only tracked satellite was visible for less than 100~s.

Each candidate arc is then assigned to one of three classes: ingress and egress arcs, in which the tangent point progressively penetrates into and emerges from the ionosphere, respectively, and tangent arcs, in which the tangent point remains within the ionosphere throughout an uninterrupted pass, without being blocked by the Earth while the signal is tracked.\\
The classification relies on two complementary criteria: the sign and magnitude of $\widehat{\mathrm{ROT}}$ compared to a data-driven threshold $\delta = 0.007$~TECU/s and the temporal location of the ROTI maximum within the arc.
The given threshold $\delta$ was derived from arcs whose ROTI series pass a robust noise screening, i.e., Tukey’s fences$^{26}$ on the Median Absolute Deviation (MAD)-based ROTI residual, excluding arcs missing more than 50\% of their samples. 
$\delta$ corresponds both to the upper bound of $|\widehat{\mathrm{ROT}}|$ for tangent arcs identified from the tangent-point geometry (median plus three MADs) and to the 10th percentile of $|\widehat{\mathrm{ROT}}|$ for ingress and egress arcs. 
In other words, $\delta$ marks the point where the near-zero $\widehat{\mathrm{ROT}}$ of a signal grazing the ionosphere ends and the steady STEC trend of a signal entering or leaving it begins. The maximum is considered to lie at the beginning or at the end of the arc if it falls within the first or last 5\% of the arc samples, respectively. For ingress and egress arcs, the arc is then trimmed so that its start or end, respectively, coincides with the ROTI minimum, as detailed below. Since trimming changes the analysed window, $\widehat{\mathrm{ROT}}$ and the ROTI extrema are recomputed on the trimmed arc and .

\subsubsection*{Ingress}
Plasma dispersive effects become increasingly evident as the signal penetrates deeper into the ionosphere, producing, in the absence of  meaningful ionospheric irregularities, a smooth, monotonic variation in the STEC time series. The ROT quantifies the magnitude and sign of this variation: a signal is flagged as a candidate ingress when $\widehat{\mathrm{ROT}}>\delta$, with a positive trend consistent with ingress. This condition is necessary but not sufficient, and is therefore complemented by the ROTI time series. The onset of a non-negligible dispersive contribution is marked by a minimum in ROTI, which defines the arc start, while the arc end coincides with the end of the segment. ROTI further discriminates between arc types through the temporal location of its maximum. In particular, for an ingress arc, the ROTI is expected to present its maximum at or near the end of the arc.

\subsubsection*{Egress}
Egress arcs are the dual of ingress arcs: the plasma effect on the signal decreases as the raypath progressively exits the ionosphere. This results in an opposite behavior of $\widehat{\mathrm{ROT}}$: an overall negative value is associated with an egress. In particular, $\widehat{\mathrm{ROT}}<-\delta$. Similarly, the ROTI behavior is mirrored with respect to the ingress arc: the maximum is located at or near the beginning of the arc. On the other hand, the timing information is assigned so that the arc start is set to the arc start epoch, while the arc end is assigned to the epoch corresponding to the ROTI minimum.

\subsubsection*{Tangent}
An arc is flagged as tangent when ingress and egress occur consecutively without loss of tracking, as when the raypath grazes the ionosphere in the absence of Earth blockage. The opposite ROT contributions of the two phases largely cancel, driving $\widehat{\mathrm{ROT}}$ towards zero. Since an exact null median is unrealistic, a tangent arc is identified by the condition $|\widehat{\mathrm{ROT}}| \leq \delta$.

Depending on how deeply the raypath grazes the ionosphere, the ROTI profile is either U-shaped or W-shaped; the deeper the penetration, the closer the profile is to a W-shaped, as the signal passes ``below'' the peak of the electron density$^{6}$. 
A U-shaped profile has boundary values larger than interior ones, whereas a W-shaped profile exhibits a central local maximum whose amplitude grows with penetration depth. Accordingly, the ROTI minimum of a tangent arc lies in its interior, while the maximum lies either at the arc boundaries (U) or at the central cusp (W). The arc start and end coincide with the first and last samples.

Single-segment arcs whose ROTI maximum lies at one of the two edges are also classified as tangent if $|\widehat{\mathrm{ROT}}| < \delta$, since a near-zero median excludes the monotonic STEC variation expected for an ingress or an egress.

Owing to the hard threshold $\delta$, the ROT and ROTI criteria may be mutually inconsistent, preventing a definitive classification. In such cases, a fallback class based solely on $\widehat{\mathrm{ROT}}$ is assigned and should be interpreted with caution, as not all classification criteria are satisfied.\\\\

Finally, a consistency check is applied across consecutive arcs. A full occultation produces a prolonged signal interruption ($T_g > 1800$\,s in this catalog) and is therefore expected to yield an ingress followed by an egress. When two consecutive arcs are classified as ingress-ingress or egress-egress, one of the two labels is necessarily incorrect, and the classification of one arc is overridden to restore the chronological ingress-egress order. Arcs modified in this way were verified against the occultation geometry, which confirmed the corrected classification.

\section*{Data Records}
The derived arc catalog is deposited in Hierarchical Data Format version 5 (HDF5) format  at the Zenodo repository  and can be accessed under the same name (https://zenodo.org/records/23216092). 
The catalog is organized into two hierarchical levels: a file-level \textit{Header}, containing dataset-wide metadata and summary statistics, and a \textit{Body}, containing one entry per identified occultation arc. The dataset structure is provided in Figure~3.

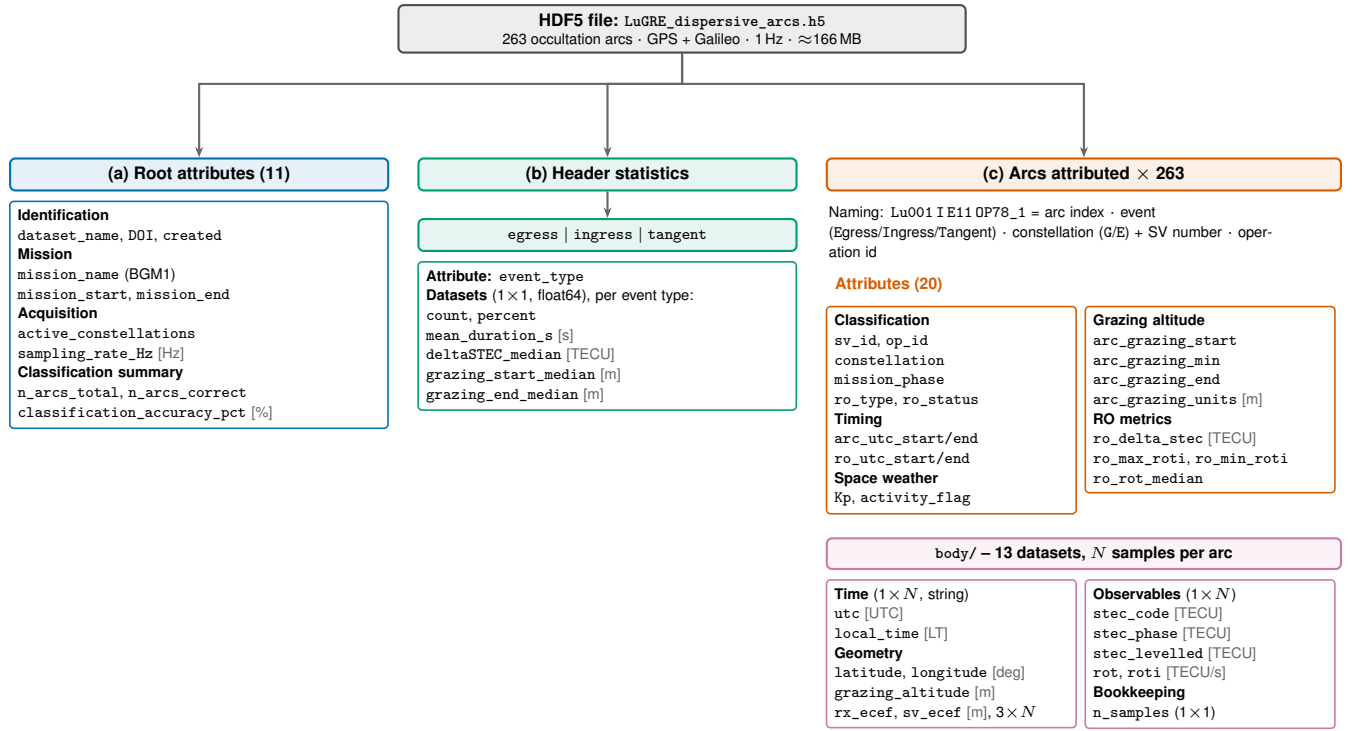
\begin{figure}
    \centering

\begin{tikzpicture}[x=1cm,y=1cm]
 
\node[grp=cRoot, minimum width=7.5cm] (root) at (8.9,0)
  {HDF5 file: \id{LuGRE\_dispersive\_arcs.h5}\\[-1pt]
   {\normalfont\fontsize{6}{7}\selectfont 263 occultation arcs $\cdot$ GPS + Galileo $\cdot$ 1\,Hz $\cdot$ $\approx$166\,MB}};
 
\node[grp=cMeta, minimum width=5.0cm, anchor=north west] (gA) at (0,-1.7) {(a) Root attributes (11)};
\node[bx=cMeta, text width=4.79cm] (bA) at ([yshift=-0.12cm]gA.south west)
{%
\textbf{Identification}\\
\id{dataset\_name}, \id{DOI}, \id{created}\\
\textbf{Mission}\\
\id{mission\_name} (BGM1)\\
\id{mission\_start}, \id{mission\_end}\\
\textbf{Acquisition}\\
\id{active\_constellations}\\
\id{sampling\_rate\_Hz} \un{Hz}\\
\textbf{Classification summary}\\
\id{n\_arcs\_total}, \id{n\_arcs\_correct}\\
\id{classification\_accuracy\_pct} \un{\%}
};
 
\node[grp=cHead, minimum width=5.0cm, anchor=north west] (gB) at (5.4,-1.7) {(b) Header statistics};
\node[grp=cHead, minimum width=5.0cm, anchor=north west,
      font=\fontsize{6.5}{7.5}\selectfont] (evt) at ([yshift=-0.35cm]gB.south west)
  {\id{egress} $|$ \id{ingress} $|$ \id{tangent}};
\node[bx=cHead, text width=4.79cm] (bB) at ([yshift=-0.12cm]evt.south west)
{%
\textbf{Attribute:} \id{event\_type}\\
\textbf{Datasets} (1$\times$1, float64), per event type:\\
\id{count}, \id{percent}\\
\id{mean\_duration\_s} \un{s}\\
\id{deltaSTEC\_median} \un{TECU}\\
\id{grazing\_start\_median} \un{m}\\
\id{grazing\_end\_median} \un{m}
};
 
\node[grp=cArc, minimum width=6.8cm, anchor=north west] (gC) at (10.8,-1.7) {(c) Arcs attributed $\times$ 263};
\node[font=\fontsize{6}{7.4}\selectfont, align=left, text width=6.6cm, anchor=north west,
      inner sep=1pt, minimum height=0.85cm] (nm) at ([yshift=-0.1cm]gC.south west)
{Naming: \id{Lu001\,I\,E11\,OP78\_1} = arc index $\cdot$ event (\id{E}gress/\id{I}ngress/\id{T}angent) $\cdot$ constellation (\id{G}/\id{E}) + SV number $\cdot$ operation id};
\node[ttl=cArc, anchor=north west] (tA) at ([yshift=-0.05cm]nm.south west) {Attributes (20)};
\node[bx=cArc, text width=3.09cm] (bC1) at ([yshift=-0.05cm]tA.south west)
{%
\textbf{Classification}\\
\id{sv\_id}, \id{op\_id}\\
\id{constellation}\\
\id{mission\_phase}\\
\id{ro\_type}, \id{ro\_status}\\
\textbf{Timing}\\
\id{arc\_utc\_start/end}\\
\id{ro\_utc\_start/end}\\
\textbf{Space weather}\\
\id{Kp}, \id{activity\_flag}
};
\node[bx=cArc, text width=3.09cm] (bC2) at ([xshift=0.1cm]bC1.north east)
{%
\textbf{Grazing altitude}\\
\id{arc\_grazing\_start}\\
\id{arc\_grazing\_min}\\
\id{arc\_grazing\_end}\\
\id{arc\_grazing\_units} \un{m}\\
\textbf{RO metrics}\\
\id{ro\_delta\_stec} \un{TECU}\\
\id{ro\_max\_roti}, \id{ro\_min\_roti}\\
\id{ro\_rot\_median}
};
 
\node[grp=cBody, minimum width=6.8cm, anchor=north west,
      font=\bfseries\fontsize{6.5}{7.5}\selectfont] (gBody) at ([yshift=-0.3cm]bC1.south west)
  {\id{body/} -- 13 datasets, $N$ samples per arc};
\node[bx=cBody, text width=3.09cm] (bD1) at ([yshift=-0.12cm]gBody.south west)
{%
\textbf{Time} (1$\times N$, string)\\
\id{utc} \un{UTC}\\
\id{local\_time} \un{LT}\\
\textbf{Geometry}\\
\id{latitude}, \id{longitude} \un{deg}\\
\id{grazing\_altitude} \un{m}\\
\id{rx\_ecef}, \id{sv\_ecef} \un{m}, 3$\times N$
};
\node[bx=cBody, text width=3.09cm] (bD2) at ([xshift=0.1cm]bD1.north east)
{%
\textbf{Observables} (1$\times N$)\\
\id{stec\_code} \un{TECU}\\
\id{stec\_phase} \un{TECU}\\
\id{stec\_levelled} \un{TECU}\\
\id{rot}, \id{roti} \un{TECU/s}\\
\textbf{Bookkeeping}\\
\id{n\_samples} (1$\times$1)
};
 
\draw[lnk] (root.south) -- ++(0,-0.4) -| (gA.north);
\draw[lnk] (root.south) -- ++(0,-0.4) -| (gB.north);
\draw[lnk] (root.south) -- ++(0,-0.4) -| (gC.north);
\draw[lnk] (gB.south) -- (evt.north);
 
\end{tikzpicture}
\caption{\textbf{Structure of the HDF5 file}. (a) Root attributes; (b) Arc statistics, with one set of statistics per arc type; (c) Arc descriptive fields, with 20 attributes per arc and the body group of 13 time series.}
\label{fig:HDF5_diagram}
\end{figure}

\subsubsection*{Arc identifier}
Each cataloged arc is uniquely identified by a composite arc identifier of the form \texttt{Lu\#\#\#@\&\&\&\$\$\$\$\$\$}, where \texttt{\#\#\#} is a three-digit unique number associated to a specific arc, \texttt{@} is the \texttt{ro\_type} code (\texttt{I}, \texttt{E} or \texttt{T}, corresponding to ingress, egress and tangent, respectively), \texttt{\&\&\&} is the \texttt{sv\_id}, and \texttt{\$\$\$\$\$\$} is the \texttt{op\_id} (e.g.,\ \texttt{Lu001IG14OP40\_0}). 
This composite identifier supplements, rather than replaces, the explicit \texttt{op\_id}, \texttt{sv\_id}, and \texttt{ro\_type} fields, which are also stored individually within each arc entry.

\subsubsection*{Header}
The file header stores dataset-wide attributes, summarized in Table~1, together with per-arc-type summary statistics (Table~2), computed independently for each arc type.

\begin{table}[t]
\footnotesize
    \centering
    \caption{File-level header attributes. }
    \begin{tabular}{lp{0.60\linewidth}}
        \toprule
        \textbf{Field} & \textbf{Description} \\
        \midrule
        \texttt{DOI} & Digital Object Identifier of the deposited dataset \\
        \texttt{active\_constellations} & GNSS constellations included (GPS, Galileo) \\
        \texttt{classification\_accuracy\_pct} & Agreement rate with respect to geometric validation \\
        \texttt{created} & Date and time of dataset generation \\
        \texttt{dataset\_name} & Identifying name of the deposited file \\
        \texttt{mission\_end} & UTC timestamp of the last mission epoch \\
        \texttt{mission\_name} & Name of the source mission (BGM1) \\
        \texttt{mission\_start} & UTC timestamp of the first mission epoch \\
        \texttt{n\_arcs\_correct} & Number of arcs in agreement with validation\\
        \texttt{n\_arcs\_total} & Total number of identified and classified arcs \\
        \texttt{sampling\_rate\_Hz} & Observable sampling rate (1~Hz) \\
        \bottomrule
    \end{tabular}
\end{table}

\begin{table}[t]
\footnotesize
    \centering
    \caption{Per-arc-type summary statistics included in the header, computed independently for each arc classes.}
    \begin{tabular}{lp{0.60\linewidth}}
        \toprule
        \textbf{Field} & \textbf{Description} \\
        \midrule
        \texttt{count} & Total number of arcs of this type \\
        \texttt{deltaSTEC\_median} & Median STEC variation across the arc \\
        \texttt{grazing\_end\_median} & Median grazing altitude at arc end \\
        \texttt{grazing\_start\_median} & Median grazing altitude at arc start \\
        \texttt{mean\_duration\_s} & Mean duration of arcs of this type \\
        \texttt{percent} & Share of arcs of this type with respect to the total \\
        \bottomrule
    \end{tabular}
\end{table}

\subsubsection*{Body}
The body of the file contains one entry per each identified occultation arc, each labelled with its composite arc identifier as described above. Each entry stores a set of scalar attributes (Table~3) and a set of associated time series (Table~4).

\begin{table}[t]
\footnotesize
    \centering
    \caption{Scalar attributes stored for each cataloged arc. }
    \begin{tabular}{lp{0.60\linewidth}}
        \toprule
        \textbf{Field} & \textbf{Description} \\
        \midrule
        \texttt{Kp} & planetary Kp index \\
        \texttt{activity\_flag} & Maximum geomagnetic activity level, following the NOAA G-scale, derived from the planetary Kp index over the arc timespan \\ 
        \texttt{arc\_grazing\_end} & Grazing altitude at arc end \\
        \texttt{arc\_grazing\_min} & Minimum grazing altitude within the arc \\
        \texttt{arc\_grazing\_start} & Grazing altitude at arc start \\
        \texttt{arc\_grazing\_units} & m \\
        \texttt{arc\_utc\_end} & UTC timestamp of arc end \\
        \texttt{arc\_utc\_start} & UTC timestamp of arc start \\
        \texttt{constellation} & GNSS constellation (GPS or Galileo) \\
        \texttt{mission\_phase} & COMMISSIONING, TRANSIT, LLO, SURFACE \\ 
        \texttt{op\_id} & Operation identifier \\
        \texttt{ro\_delta\_stec} & STEC variation associated with the arc \\
        \texttt{ro\_max\_roti} & Maximum ROTI value within the arc \\
        \texttt{ro\_min\_roti} & Minimum ROTI value within the arc \\
        \texttt{ro\_rot\_median} & Median Rate of TEC over the arc \\
        \texttt{ro\_status} & Validation outcome (correct/wrong) with respect to geometric validation  \\
        \texttt{ro\_type} & Assigned arc class (ingress, egress, tangent) \\
        \texttt{ro\_utc\_end} & UTC timestamp end of entire radio occultation \\
        \texttt{ro\_utc\_start} & UTC timestamp start of entire radio occultation\\
        \texttt{sv\_id} & Satellite vehicle identifier \\
        \bottomrule
    \end{tabular}
\end{table}

\begin{table}[t]
\footnotesize
    \centering
    \caption{Time series stored for each cataloged arc.}
    \begin{tabular}{lp{0.60\linewidth}}
        \toprule
        \textbf{Field} & \textbf{Description} \\
        \midrule
        \texttt{grazing\_altitude} & Grazing altitude in Earth Centered Earth Fixed (ECEF) reference frame \\
        \texttt{latitude} & Sub-tangent-point latitude \\
        \texttt{local\_time} & Time series of the tangent point local time \\
        \texttt{longitude} & Sub-tangent-point longitude \\
        \texttt{n\_samples} & Number of samples in the arc \\
        \texttt{rot} & Rate of TEC over the tracking arc \\
        \texttt{roti} & Rate of TEC Index over the tracking arc \\
        \texttt{rx\_ecef} & LuGRE receiver position, ECEF reference frame \\
        \texttt{stec\_code} & Code-derived STEC (Eq.~1) \\
        \texttt{stec\_levelled} & Carrier-phase-leveled STEC (Eq.~5) \\
        \texttt{stec\_phase} & Phase-derived STEC (Eq.~2) \\
        \texttt{sv\_ecef} & GNSS satellite position, ECEF reference frame \\
        \texttt{utc} & UTC timestamp of each sample \\
        \bottomrule
    \end{tabular}
\end{table}

It should be noted that the stored time series are \textbf{not} restricted to the identified arc timespan (i.e., \texttt{arc\_utc\_start}, \texttt{arc\_utc\_end}) neither associated to the arc-validity mask, but instead span the full tracking arc (i.e., \texttt{ro\_utc\_start}, \texttt{ro\_utc\_end}), i.e.,\ the continuous interval over which the satellite was tracked. Therefore, it is possible that all those double-frequency based time series presents some natural gaps whenever one of the two frequencies data record is missing.\\\\
An example of information extracted from the dataset is reported in Figure~4.

\begin{figure*}[t]
    \centering
        \includegraphics[trim={0cm 1cm 0cm 0cm},clip,width=1\linewidth]{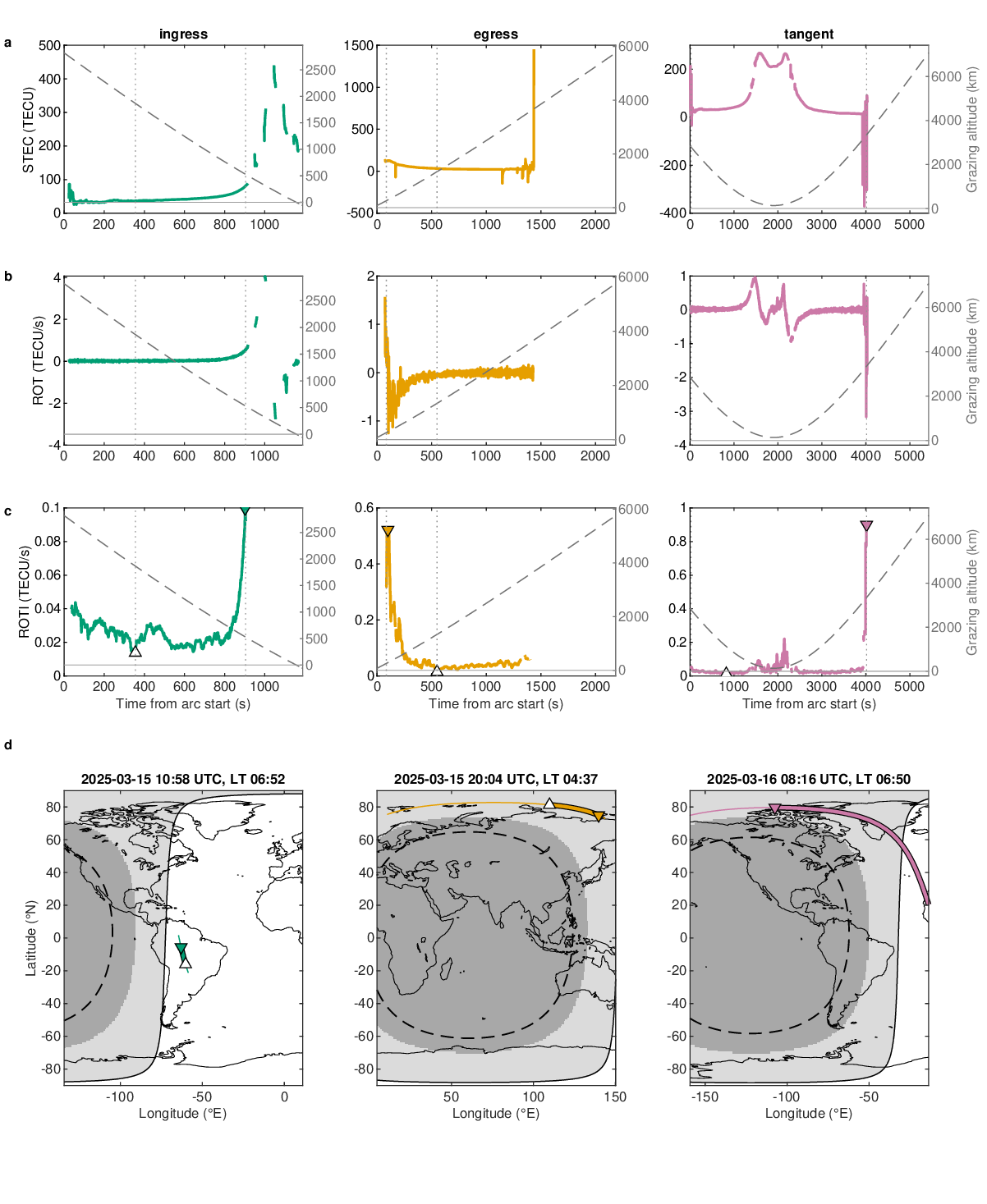}
        \caption{Columns show one ingress (green), one egress (orange) and one tangent (purple) arc (identifiers: \texttt{Lu190IG18OP74\_0}, \texttt{Lu229EE36OP76\_0}, \texttt{Lu240TG11OP76\_0}). \textbf{a--c,} Leveled STEC (\textbf{a}), ROT (\textbf{b}) and ROTI (\textbf{c}) as a function of time from arc start. The grey solid line (right axis) is the grazing altitude of the tangent point; the horizontal grey line marks the surface (0~km). Dotted vertical lines delimit the timespan of the occultation arc. Triangles in \textbf{c} mark the ROTI maximum ($\blacktriangledown$) and minimum ($\triangle$). \textbf{d,} Ground track of the tangent point. The thick line represent the arc, with start and end marked by a the same markers as panel \textbf{c}. Shading shows day (white), twilight (light grey; solar zenith angle $90-108^\circ$) and night (dark grey) at the mid-point of the arc; the solid and dashed lines are the terminators at the surface and at the median grazing altitude of the tangent point, respectively. Panel titles give UTC and local time at the same instant.}
    \label{fig:sample_plots}
\end{figure*}

Table~5 summarises the outcome of the classification. For each arc type, it reports the number of arcs, their percentage within the corresponding group, and their mean duration, both for the whole catalog and broken down by constellation and mission phase.

\begin{table}[!htb]
\footnotesize
\centering
\setlength{\tabcolsep}{1.5pt}
\caption{Classification statistics: number of arcs ($n$), percentage with respect to the corresponding group (\%), and mean arc duration ($\bar{t}$), broken down by arc type, constellation, and mission phase.}
\label{tab:classStats}
\begin{tabular}{l
                S[table-format=3] S[table-format=2.1] S[table-format=4]
                @{\hspace{3pt}\vrule\hspace{3pt}}
                S[table-format=3] S[table-format=2.1] S[table-format=4]
                S[table-format=3] S[table-format=2.1] S[table-format=4]
                @{\hspace{3pt}\vrule\hspace{3pt}}
                S[table-format=3] S[table-format=2.1] S[table-format=4]
                S[table-format=3] S[table-format=2.1] S[table-format=4]
                S[table-format=3] S[table-format=2.1] S[table-format=4]
                S[table-format=3] S[table-format=2.1] S[table-format=4]}
\toprule
 & \multicolumn{3}{c}{\textbf{Overall}}
 & \multicolumn{6}{c}{\textbf{Constellation}}
 & \multicolumn{12}{c}{\textbf{Mission phase}} \\
 & \multicolumn{3}{c}{}
 & \multicolumn{3}{c}{\textit{Galileo}}
 & \multicolumn{3}{c}{\textit{GPS}}
 & \multicolumn{3}{c}{\textit{Comm.}}
 & \multicolumn{3}{c}{\textit{Transit}}
 & \multicolumn{3}{c}{\textit{Low Lunar Orbit}}
 & \multicolumn{3}{c}{\textit{Surface}} \\
\cmidrule(lr){2-4}
\cmidrule(lr){5-7}\cmidrule(lr){8-10}
\cmidrule(lr){11-13}\cmidrule(lr){14-16}\cmidrule(lr){17-19}\cmidrule(lr){20-22}
\textbf{\texttt{ro\_type}}
 & {$n$} & {\%} & {$\bar{t}$\,[s]}
 & {$n$} & {\%} & {$\bar{t}$\,[s]}
 & {$n$} & {\%} & {$\bar{t}$\,[s]}
 & {$n$} & {\%} & {$\bar{t}$\,[s]}
 & {$n$} & {\%} & {$\bar{t}$\,[s]}
 & {$n$} & {\%} & {$\bar{t}$\,[s]}
 & {$n$} & {\%} & {$\bar{t}$\,[s]} \\
\midrule
{Total}
 & {263} & {100}  & {--}
 & {162} & {61.6} & {--}
 & {101} & {38.4} & {--}
 & {15}  & {5.7}  & {--}
 & {21}  & {8.0}  & {--}
 & {8}   & {3.0}  & {--}
 & {219} & {83.3} & {--} \\
\midrule
Egress
 & 94  & 35.7 & 367
 & 59  & 36.4 & 348
 & 35  & 34.7 & 398
 & 3   & 20.0 & 423
 & 9   & 42.9 & 199
 & 3   & 37.5 & 348
 & 79  & 36.1 & 384 \\
Ingress
 & 89  & 33.8 & 450
 & 58  & 35.8 & 433
 & 31  & 30.7 & 480
 & 1   & 6.7  & 551
 & 8   & 38.1 & 304
 & 5   & 62.5 & 648
 & 75  & 34.2 & 451 \\
Tangent
 & 80  & 30.4 & 2110
 & 45  & 27.8 & 2105
 & 35  & 34.7 & 2116
 & 11  & 73.3 & 1068
 & 4   & 19.0 & 792
 & {--}& {--} & {--}
 & 65  & 29.7 & 2367 \\
\bottomrule
\end{tabular}
\end{table}

\section*{Technical Validation}

The classification validation is performed after the entire identification/classification process by adopting an approach solely based on the geometry and tangent point dynamics.
The receiver position, which can be obtained from the LuGRE data ground truth$^{13}$
, is used to recover the signal time of flight and therefore the transmission time with respect to a specific satellite position obtained by processing GNSS satellite ephemeris files$^{3}$.  
For operations spanning two consecutive days, the broadcast ephemerides of the start day are used for the whole operation.
Since the Earth rotates during the signal propagation, it is necessary to  take into account this effect, which introduces a propagation error on the observables. To compensate this non-ideality, the Sagnac correction is applied to the ECEF satellite positions$^{3}$. 
Subsequently, the GNSS satellite to the receiver geometric line of sight vector is then obtained as $\mathbf{r}_{\mathrm{LOS}} = \mathbf{r}_{\mathrm{GNSS}} - \mathbf{r}_{\mathrm{rx}}$
where $\mathbf{r}_{\mathrm{GNSS}}$ and and $\mathbf{r}_{\mathrm{rx}}$ are the satellite position at transmission time and the receiver position at reception time, respectively. The point of closest ray's approach to the Earth's center, namely the tangent point, is given by 
\begin{equation}\label{eq:tangent_point}
    \mathbf{p} = \mathbf{r}_{\mathrm{rx}} - 
    \left(\mathbf{r}_{\mathrm{rx}} \cdot \hat{\mathbf{r}}_{\mathrm{LOS}}\right)
    \hat{\mathbf{r}}_{\mathrm{LOS}},
\end{equation}
where $\hat{\mathbf{r}}_{\mathrm{LOS}} = \mathbf{r}_{\mathrm{LOS}} / \|\mathbf{r}_{\mathrm{LOS}}\|_2$ is the unit vector. To extract the grazing altitude is necessary to convert $\mathbf{p}$ from ECEF coordinates to geodetic latitude, longitude, and altitude using the World Geodetic System 1984 (WGS84) reference ellipsoid$^{3}$. 
From the epoch-by-epoch analysis of the grazing altitude it is possible to extract a direct indication of which dispersive layers are being probed by the GNSS signal, and is computed as
\begin{equation}
    h = \rho\cos\varphi
      + \bigl(p_{z} + e^{2}\,N\sin\varphi\bigr)\sin\varphi
      - N ,
    \label{eq:h_ellipsoidal}
\end{equation}
where,
\begin{equation}
    N =\frac{a}{\sqrt{1 - e^2 \sin^2\!\varphi}},
\end{equation}
$e$ is the first eccentricity of the reference ellipsoid, $a$ is the semi-major axis of the reference ellipsoid,$\rho =  \sqrt{p_{x}^2 + p_{y}^2}$ is the radial distance from the polar axis obtained from $p_{x}$ and $p_{y}$ tangent point ECEF coordinates, $\varphi$ is the geodetic latitude and finally, $p_{z}$ is the signed distance from the equatorial plane$^{3}$. 
Note that the deviation of the actual propagation path from the straight line of sight used to define $\hat{\mathbf{r}}_{\mathrm{LOS}}$ and the tangent point  $\mathbf{p}$,i.e. ionospheric ray bending, is neglected in this geometric validation. However for the classification is not critical as this approximation can introduce horizontal and vertical shifts of the true grazing point on the order of a few kilometers$^{27}$
, which are small compared to the spatial scales relevant for the present classification and are therefore considered negligible.

At this point, the temporal evolution of the grazing altitude hinders the nature of each arc. Relying on geometry-based criteria is possible to perform a classification i.e. comparing the median of the grazing altitude time derivative $\mathrm{\widehat{v_h}}$ against a data-driven threshold $\gamma$ = 1000 m/s set according to the fastest tangent arc.

\textbf{Ingress:} the first altitude estimate exceeds the last one and the minimum corresponds to the last sample in the series as the tangent point gets closer to the Earth horizon. Furthermore, since the grazing altitude decreases $\mathrm{\widehat{v_h}} < -\gamma$.

\textbf{Egress:} on the other hand, when the tangent point leaves the Earth horizon, the initial altitude results lower than the final one and the minimum corresponds to the first sample in the series. Consequently the last estimate is the highest point of the arc. Correspondingly, $\mathrm{\widehat{v_h}} > \gamma$.

\textbf{Tangent:} the grazing altitude minimum is located at the interior of the series. Both the initial and final altitudes exceed the minimum.

Comparing the outcomes of the geometric approach and the results of the proposed method is possible to spot whether there are some classifications that deserve specific analyses.


\section*{Usage Notes}

Arc classification is based on the dispersive effects induced by TEC and validated through the geometry of the tangent point. Arcs generated by sidelobe transmissions and associated with high grazing altitudes may be affected by dispersive effects induced primarily by the plasmasphere, and only to a lesser extent by the upper layer ionosphere. The distinction between mainlobe and sidelobe arcs is out of the scope of this catalog. However, since sidelobe observations are poorly characterized in the literature, they may lead to misleading classifications and should be treated with caution.

The geometric validation described in the Technical Validation section disagrees with the classification for 30 of the 263 arcs, corresponding to a disagreement rate of 11.4\%. Users should be aware of the main causes of such disagreements. First, some arcs are truncated by the boundaries of the observation windows: in these cases, the $\widehat{\mathrm{ROT}}$ computed over a partial arc may not be representative, and the ROTI, computed over a 30-s sliding window, loses samples near the arc edges and becomes poorly informative. Second, whenever the ROT and ROTI features are inconsistent, the classification relies only on the thresholded $\widehat{\mathrm{ROT}}$, which is sensitive to such edge effects and to other non-idealities. Third, gaps caused by interruptions of dual-frequency tracking make the ROT and ROTI series intermittent and harder to interpret. Arcs whose classification disagrees with the geometric validation are flagged in the \texttt{ro\_status} attribute, and we recommend inspecting, or excluding them, depending on the intended application.



\section*{Data Availability}
The \texttt{LuGRE radio occultation catalog} recorded in HDF5 format and presented through this article has been deposited in its version v1.0.0 on the Zenodo online repository and can be accessed under the same name (https://zenodo.org/records/23216092). The dataset repository is linked to the Cislunar Earth's Ionosphere and Plasmasphere Observation Network (CLIP-ON) Zenodo community (https://zenodo.org/communities/clip-on/), aiming at collecting space climatology contributions derived from the LuGRE data and future missions observations from the cislunar volume. 

\section*{Code Availability}
The code used for the creation of the dataset can be found at https://zenodo.org/records/23330533. The code can be run in MATLAB on both macOS/Linux and Windows platforms. In particular, the code was written in MATLAB 2022b, so this version is suggested to avoid any compatibility problems.
A README file describing the code usage is present in the same repository.

\section*{Author Contributions}
A.M. and C.C. conceived the catalog concept. F.F. and A.M. developed the dataset structure that was then refined and approved by all the other authors. F.F. designed the experiment, the analytical and validation strategy and the code. R.G. and M.G. contributed to the catalog by adding the local time time series, Kp index and G-Scale evaluation attributes for each arc. F.F. and A.M. wrote the manuscript. F.F., A.M. and A.N. submitted the dataset and the code for open access. L.S., C.C. and D.F. (LuGRE co-PI) supervised the study. All the authors contributed to the revision of the manuscript.

\section*{Competing Interests}
The authors declare no competing interest.

\section*{Funding}
The creation of this dataset was funded within the contract n. 2021-26-HH.0 between Agenzia Spaziale Italiana and Politecnico di Torino ”Attività di Ricerca e Sviluppo inerente alla Navigazione GNSS nello Space volume Terra/Luna nell’ambito del Lunar GNSS Receiver Experiment”.


\section*{References}
\begin{enumerate}



\item Kelley, M. C. \textit{The Earth's Ionosphere: Plasma Physics and Electrodynamics} 2nd edn, Vol. 96 of International Geophysics Series (Academic Press, 2009).

\item Smirnov, A. \textit{et al.} Bridging the gap between the Earth's ionosphere and plasmasphere. \textit{Geophys. Res. Lett.} \textbf{52}, e2025GL119267 (2025). https://doi.org/10.1029/2025GL119267 

\item Teunissen, P. J. G. \& Montenbruck, O. (eds.) \textit{Springer Handbook of Global Navigation Satellite Systems} (Springer, 2017). https://doi.org/10.1007/978-3-319-42928-1 

\item Darrouzet, F. \textit{et al.} Plasmaspheric density structures and dynamics: properties observed by the CLUSTER and IMAGE missions. \textit{Space Sci. Rev.} \textbf{145}, 55--106 (2009). https://doi.org/10.1007/s11214-008-9438-9 

\item Cherniak, I. V., Zakharenkova, I. E., Krankowski, A. \& Shagimuratov, I. I. Plasmaspheric electron content derived from GPS TEC and FORMOSAT-3/COSMIC measurements: solar minimum condition. \textit{Adv. Space Res.} \textbf{50}, 427--440 (2012). https://doi.org/10.1016/j.asr.2012.04.002 

\item Cesaroni, C., Spogli, L., Guerra, M., Ghidoni, R., Ventriglia, V., Minetto, A., Nardin, A., Alfonsi, L., and Dovis, F. (2026). Observing the earth’s plasmasphere and ionosphere from the lunar surface. Geophysical Research Letters, 53(12): https://doi.org/10.1029/2026GL121811. 

\item Jakowski, N. \textit{et al.} GPS radio occultation measurements of the ionosphere from CHAMP: early results. \textit{Geophys. Res. Lett.} \textbf{29}, 95-1--95-4 (2002). https://doi.org/10.1029/2001GL014364 

\item NOAA Space Weather Prediction Center. Solar cycle progression: F10.7 cm radio flux. \textit{NOAA/SWPC} \url{https://www.swpc.noaa.gov/products/solar-cycle-progression} (2026). Accessed 9 June 2026. 

\item Tapping KF. 2013. The 10.7 cm solar radio flux (F10. 7). Space Weather 11(7): 394–406. https://doi.org/10.1002/swe.20064 

\item NOAA Space Weather Prediction Center. NOAA space weather scales. \textit{NOAA/SWPC} \url{https://www.swpc.noaa.gov/noaa-scales-explanation} (2026). Accessed 9 June 2026. 

\item Matzka, J., Stolle, C., Yamazaki, Y., Bronkalla, O. \& Morschhauser, A. The geomagnetic Kp index and derived indices of geomagnetic activity. \textit{Space Weather} \textbf{19}, e2020SW002641 (2021). https://doi.org/10.1029/2020SW002641 

\item Parker, J. J. K. \textit{et al.} GNSS reception at the Moon: first results of the Lunar GNSS Receiver Experiment (LuGRE). \textit{NAVIGATION} \textbf{73}, navi.756 (2026). https://doi.org/10.33012/navi.756 

\item Parker, J. J. K. \textit{et al.} Lunar GNSS Receiver Experiment (LuGRE) mission data. \textit{Zenodo} \url{https://doi.org/10.5281/zenodo.16411687} (2025).

\item Minetto, A., Fiorina, F. \textit{et al.}Identification of Dispersive GNSS Occultation Arcs Observed by LuGRE Along the Blue Ghost Mission 1 Trajectory. In \textit{Proc. 39th International Technical Meeting of the Satellite Division of The Institute of Navigation (ION GNSS+ 2026)} 1457--1471 (Institute of Navigation, 2026). https://doi.org/10.33012/2026.20747 

\item Acton, C. H. Ancillary data services of NASA's Navigation and Ancillary Information Facility. \textit{Planet. Space Sci.} \textbf{44}, 65--70 (1996). https://doi.org/10.1016/0032-0633(95)00107-7 

\item Noll, C. E. The Crustal Dynamics Data Information System: a resource to support scientific analysis using space geodesy. \textit{Adv. Space Res.} \textbf{45}, 1421--1440 (2010). https://doi.org/10.1016/j.asr.2010.01.018 

\item Matzka, J., Bronkalla, O., Tornow, K., Elger, K. \& Stolle, C. Geomagnetic Kp index. V. 1.0. \textit{GFZ Data Services} https://doi.org/10.5880/Kp.0001 (2021). 

\item Minetto, A. \textit{et al.} LuNART-q: the LuGRE quick navigation analysis and reporting tool. In \textit{Proc. 38th International Technical Meeting of the Satellite Division of The Institute of Navigation (ION GNSS+ 2025)} 1377--1390 (2025). https://doi.org/10.33012/2025.20440 

\item Grubbs, F. E. (1969). Procedures for Detecting Outlying Observations in Samples. Technometrics, 11(1), 1–21. https://doi.org/10.1080/00401706.1969.10490657 

\item Weber, J. Phase, group, and signal velocity. \textit{Am. J. Phys.} \textbf{22}, 618--620 (1954). \url{https://doi.org/10.1119/1.1933858}

\item Skone, S. \textit{Wide Area Ionosphere Grid Modeling in the Auroral Region}. Ph.D. thesis, Department of Geomatics Engineering, The University of Calgary (1998)

\item Pi, X., Mannucci, A. J., Lindqwister, U. J. \& Ho, C. M. Monitoring of global ionospheric irregularities using the worldwide GPS network. \textit{Geophys. Res. Lett.} \textbf{24}, 2283--2286 (1997). https://doi.org/10.1029/97GL02273 

\item Luo, X., Gu, S., Lou, Y., Cai, L. \& Liu, Z. Amplitude scintillation index derived from C/N0 measurements released by common geodetic GNSS receivers operating at 1 Hz. \textit{J. Geodesy} \textbf{94}, 27 (2020). https://doi.org/10.1007/s00190-020-01359-7

\item Kervalishvili, G., Xiong, C., Stolle, C., Rauberg, J. \& van den IJssel, J. \textit{Computation of TEC and Rate of TEC Index (ROTI) from GOCE GPS}. Report No. GO-TN-HPF-GS-0337 (GFZ German Research Centre for Geosciences / Delft University of Technology, 2018).

\item Akima, H. A new method of interpolation and smooth curve fitting based on local procedures. \textit{J. ACM} \textbf{17}, 589--602 (1970). https://doi.org/10.1145/321607.321609 

\item Tukey, J. W. \textit{Exploratory Data Analysis} (Addison-Wesley, 1977).

\item Iiyama, K. and Gao, G. (2025). Ionospheric and plasmaspheric delay characterization for lunar terrestrial gnss receivers with global core plasma model. arXiv preprint, arXiv:2510.10059. arXiv:2510.10059.  
Iiyama, K. and Gao, G. \textit{Ionospheric and plasmaspheric delay characterization for lunar terrestrial GNSS receivers with global core plasma model}. NAVIGATION: Journal of the Institute of Navigation (2025). 
https://doi.org/10.48550/arXiv.2510.10059 

\end{enumerate}

\end{document}